\documentclass[twocolumn,prl,amssymb,showpacs]{revtex4}
\usepackage{amsmath,color,graphicx}

\newcommand{\ket}[1]{\mbox{$| #1 \rangle$}}

\newcommand{\proj}[1]{\mbox{$| #1 \rangle\!\langle #1 |$}}
\def\H{\mathcal{H}}
\def\A{\mathcal{A}}
\def\B{\mathcal{B}}
\def\N{\mathcal{N}}
\def\s{\mathcal{S}^+}

\def\t{^{\mbox{\tiny T}}}
\def\c{\mathcal{C}}
\def\nv{\bf{0}}
\def\tr{\mbox{tr}}
\def\co{\mbox{conv}}
\def\o{\!\otimes\!}
\def\id{\mathbb{I}}
\newcommand{\C}{{\mathbb{C}}}
\def\vs{\vspace{.2cm}}

\begin{document}

\title{ All bipartite entangled states display some hidden  nonlocality}

\author{Llu\'{\i}s~Masanes} \affiliation{School of Mathematics, University of
  Bristol, Bristol BS8 1TW, U.K.}

\author{Yeong-Cherng~Liang} \affiliation{School of Physical Sciences, The
  University of Queensland, Queensland 4072, Australia.}

\author{Andrew~C.~Doherty} \affiliation{School of Physical Sciences, The
  University of Queensland, Queensland 4072, Australia.}

\date{\today}

\begin{abstract}
  We show that a violation of the Clauser-Horne-Shimony-Holt (CHSH)
  inequality can be demonstrated in a certain kind of Bell experiment
  if and only if the state is entangled. Our protocol allows local
  filtering measurements and involves shared ancilla states that do
  not themselves violate CHSH. Our result follows from two main
  steps. We first provide a simple characterization of the states that
  violate the CHSH-inequality after local filtering operations in terms of
  witness-like operators. Second, we prove that for each entangled
  state $\sigma$, there exists another state $\rho$ not violating CHSH,
  such that $\rho\otimes\sigma$ violates CHSH. Hence, in this
  scenario, $\sigma$ cannot be substituted by classical correlations
  without changing the statistics of the experiment; we say that
  $\sigma$ is not {\it simulable} by classical correlations and our
  result is that entanglement is equivalent to non-simulability.
\end{abstract}

\maketitle

In 1964 Bell ruled out the possibility that a local-realistic theory
could reproduce all the experimental predictions given by Quantum
Mechanics \cite{Be}. In a local-realistic theory the outcomes of
measurements are determined in advance by unknown (or hidden)
variables, and do not depend on the choice of measurements made by
distant observers. Bell's theorem states that the quantum mechanical
probabilities for outcomes of measurements distributed in space
cannot, in general, be replicated in {\it any} local-realistic
theory. This fact is demonstrated for particular states and
measurements by the violation of a Bell inequality.  Entanglement is
in some way responsible for this phenomenon since entangled states are
required to demonstrate the violation of Bell inequalities. However
since it has been known for some time that there are entangled states
that do not violate any Bell inequality~\cite{W,B} the precise
relationship between entanglement and Bell inequality violation has
remained poorly understood.

The definition of entangled state is made in terms of the physical
resources needed for the {\em preparation} of the state: a
multipartite state is said to be entangled if it cannot be prepared
from classical correlations using local quantum operations \cite{W}.
But this definition tells us nothing about the ``behavior'' of the
state. For example, does the state violate a Bell inequality, or is it
useful in some quantum protocol such as teleportation?

It is known that every pure entangled state violates a Bell inequality
\cite{Gi,Po} and that no separable state does~\cite{W}, but the
situation gets more complicated for  mixed entangled states.  There are
bipartite mixed states that, though being entangled, possess a local
hidden variables model (LHVM) whenever measurements are made on a single copy
of the state (see for example~\cite{B,W}). But some of these states
violate Bell inequalities if, prior to the measurement, the state is
processed by local operations and classical communication (LOCC)
\cite{SP,G}. Moreover, by jointly measuring more than one copy of
these states after some LOCC preprocessing, it was shown that an even
larger set of entangled states could be detected through their
violation of a Bell
inequality~\cite{P}.

Generalizing this idea one can get a strong test of the nonlocality
``hidden'' in a state by combining local filtering operations and
collective measurements: Perform joint local filtering operations on
an arbitrarily large number of copies of the state and then a Bell
inequality test on the resulting state. If the resulting probabilities
violate a Bell inequality, we say that the original state violates
this inequality asymptotically. In Ref.~\cite{Masymp} it is shown that
a bipartite state violates the Clauser-Horne-Shimony-Holt (CHSH)
inequality~\cite{CHSH} asymptotically if, and only if, it is
distillable. This result suggests that undistillable entangled states
may admit a LHVM description even when experiments are performed on
many copies of the state.

Given these negative results, it seems necessary to allow still
more general protocols for the nonlocality ``hidden" in arbitrary
entangled states to manifest itself.  One natural possibility is to
allow joint processing with auxiliary states (that do not themselves
violate the Bell inequality) rather than just with more copies of the
state in question.  This idea has been fruitful to show that useful
entanglement can be extracted from all non-separable states
\cite{Mtele} and in this Letter we use it to show that there is indeed
a kind of hidden nonlocality that is possessed by all entangled
states.

In order to investigate the possibilities of this more general kind of
hidden nonlocality we introduce the concept of a {\em simulable}
state. We say that a bipartite state $\sigma$ is simulable by
classical correlations, or just {\em simulable}, if in any protocol
(possibly involving other resources such as shared quantum states) two
separated parties sharing classical correlations instead of $\sigma$
can obtain the same statistics for the outcomes of the protocol. In
this sense, simulable states have a completely classical ``behavior".
Of course we are most interested in this Letter in the case where the
protocol concerned is a test of nonlocality. Clearly, all separable
states are simulable.  A possible way to simulate a separable state is
by just preparing it from classical correlations \cite{W}.

\vs The scenario that we consider is the typical Bell-like experiment,
where two parties share a bipartite system and perform local
measurements on it. Alice chooses between the observables $x=0,1$ and
obtains the outcomes $a=0,1$, and analogously for Bob, $y$ and $b$.
All the relevant experimental information is contained in the joint probability
distribution for the outcomes conditioned on the choice of observables
$P(a,b|x,y)$. It is convenient to define the correlation functions
\begin{equation}\label{correlators}
    C_{xy}\equiv P(a=b|x,y)-P(a\neq b|x,y)\ .
\end{equation}
By local relabeling of $(a,b,x,y)$ it is always possible to make
$C_{00}, C_{01}, C_{10} \geq 0$. With this convention, the
distribution $P(a,b|x,y)$ admits a LHVM if~\cite{A.Fine:PRL:1982}, and
only if, it satisfies the CHSH-inequality \cite{CHSH}
\begin{equation}\label{chsh}
    C_{00} +C_{01} +C_{10} -C_{11} \leq 2\ .
\end{equation}
Let us characterize the set of bipartite states that violate the
CHSH-inequality after preprocessing.

\vs {\bf Definition.} Denote by $\c$ the set of bipartite states that
do not violate the CHSH-inequality, even after stochastic local operations
without communication.

\vs By stochastic we mean that the operation can fail, and we do not
care about the probability of failure, as long as it is strictly
smaller than one. Up to normalization, these operations allow the
transformations
\begin{equation}
\label{sepmap}
  \rho
\rightarrow \Omega(\rho) = \sum_i \left(A_i\otimes B_i \right) \rho \left(A_i\otimes
  B_i \right)^\dagger,
\end{equation}
where $A_i$ and $B_i$ are, respectively, Kraus operators acting on the
first and second system. This class of maps is known as the {\it
  separable maps}.

In Ref.~\cite{Masymp} it is shown that the states in $\c$ do not violate
CHSH even after stochastic local operations {\em with} communication. So the
exact nature of the local operations allowed in the definition of $\c$
is not important.  Clearly,  states that do not violate the
CHSH-inequality asymptotically are in $\c$. Thus, $\c$ contains all
undistillable states~\cite{Masymp}.  We are now able to state
precisely the central result of this Letter.

\vs {\bf Theorem.} A bipartite state $\sigma$ is entangled if, and
only if, there exists a state $\rho \in \c$ such that
$\rho\otimes\sigma$ is not in $\c$.\vs

The consequences of this theorem are dramatic. If $\rho$ belongs to
$\c$, no matter how much additional classical correlation
(which can always be represented by a separable state
$\eta_{\mbox{\scriptsize sep}}$) we supply to it, the result $\rho
\otimes \eta_{\mbox{\scriptsize sep}}$ is still in $\c$. Contrary, the
state $\rho \otimes \sigma$ is not in $\c$ even if both $\rho$  and
$\sigma$ are in $\c$. The violation of CHSH manifests the
qualitatively different behavior between $\rho \otimes \sigma$ and
$\rho \otimes \eta_{\mbox{\scriptsize sep}}$, where
$\eta_{\mbox{\scriptsize sep}}$ is any separable state, and $\sigma$
is any entangled state. Summarizing, for each entangled state $\sigma$
there exists a protocol (which also involves the auxiliary state
$\rho$ associated with the theorem) in which $\sigma$ cannot be
substituted by an arbitrarily large amount of classical correlations
without changing the result: {\em Entangled states are the ones that
  cannot be simulated by classical correlations.}\vs

The proof of the above theorem has two main ingredients. Firstly we
note that $\c$ is a convex set and provide a characterization of $\c$
in terms of witness-like operators that detect CHSH-violation.
Secondly we use convexity arguments similar to those in Ref.~\cite{Mtele}
to prove by contradiction that there exists some $\rho\in \c$ such
that one of these witnesses may be constructed for $\rho\otimes
\sigma$ whenever $\sigma$ is entangled. To carry this argument through
we require a characterization of the separable completely positive
maps between Bell diagonal states that will be explained further
elsewhere~\cite{DLM}. Firstly we describe the witnesses for
CHSH-violation.

\vs {\bf Lemma 1.} A bipartite state $\rho$ acting on
$\H_{\A}\otimes\H_{\B}$ belongs to $\c$ if, and only if, it satisfies
\begin{equation}\label{C}
    \tr\!\left[\rho\, (A\otimes B) H_\theta
    (A\otimes B)^\dag \right]
    \geq 0\ ,
\end{equation}
for all matrices of the form $A\! :\C^2\rightarrow\H_{\A}$, $B\!
:\C^2\rightarrow\H_{\B}$ and all numbers $\theta \in [0,\pi/4]$, where
\begin{equation} \label{H}
    H_\theta\equiv \id\otimes\id-\cos\theta\,
    \sigma_x\otimes\sigma_x - \sin\theta\,
    \sigma_z\otimes \sigma_z,
\end{equation}
$\id$ being the $2\times 2$ identity matrix and
$\{\sigma_i\}_{i=x,y,z}$ the Pauli matrices.

\vs {\em Proof.} To start off, we recall that, without preprocessing,
a two-qubit state $\varrho$ violates the CHSH inequality iff
$\mu_1^2+\mu_2^2 > 1$, where $\mu_1$ and $\mu_2$ are the two largest
singular values of the $3\times 3$ real matrix $R_{ij}= \tr[\varrho\,
\sigma_i\otimes \sigma_j]$, with indices
$i,j=x,y,z$~\cite{VW,Horodecki}. Equivalently, $(\mu_1,\mu_2)$ derived
from $\varrho$ must lie outside the unit circle $\mu_1^2+\mu_2^2=1$,
which is true if and only if there exists $\theta\in [0,2\pi]$ such
that
\begin{equation}\label{Eq:Linear:CHSH}
\mu_1\cos\theta+\mu_2\sin\theta >1.
\end{equation}
Now, it is also well known that by appropriate local unitary
transformations $U$, $V$, it is always possible to arrive at a local
basis such that $R$ is diagonal with $\mu_1=R_{xx}$ and
$\mu_2=R_{zz}$. From the definition of $R$ it follows that
\begin{equation}
  \mu_1\cos\theta = \tr\!\left[(U \otimes V)
    \varrho\, (U \otimes V)^\dag (\cos\theta\,
    \sigma_x\otimes\sigma_x)
    \right],
\end{equation}
with the expression for $\mu_2 \sin \theta$ involving obvious
modifications.  Since singular values are non-negative, it thus follows
that if $\varrho$ violates CHSH then there exist $U,V\in\rm SU(2)$,
$\theta\in [0,\pi/4]$ such that
\begin{equation}\label{Cqubits}
    \tr\!\left[\varrho\, (U \otimes V)^\dag H_\theta
    (U \otimes V) \right]
    < 0\ .
\end{equation}
On the other hand suppose that there exists some $(U,V,\theta)$
satisfying (\ref{Cqubits}). Thus we have $R_{xx}\cos \theta
+R_{zz}\sin \theta > 1$. If we assume $\mu_1 \geq \mu_2$, the inequalities $|R_{xx}|,|R_{zz}| \leq
\mu_1 \leq 1$ follow from the definition of singular values and the well
known fact that all singular values of $R$ are less than one. Since $0
\leq \theta \leq \pi/4$ both $R_{xx}$ and $R_{zz}$ must be positive
and since
$\cos \theta \geq \sin \theta$ we may assume without loss of
generality that $R_{xx}\geq R_{zz}$.
The singular values of $R$ obey the inequality $|R_{xx}+R_{zz}|\leq
\mu_1+\mu_2$ \cite{Bhat} and as a result we find
$ \mu_1\cos\theta+\mu_2\sin\theta > 1$
so $\rho$ violates the CHSH-inequality.
Thus $\rho$ violates CHSH iff (\ref{Cqubits}) holds.

Let us come back to the question of CHSH violation after local
filtering operations. Assume that $\rho$ violates CHSH
after stochastic local operations. Let us show that it must
violate (\ref{C}) for some $(A,B,\theta)$. In Ref.~\cite{Masymp} it is
proven that, if a state violates CHSH then it can be transformed by
stochastic local operations into a two-qubit state which also violates
CHSH. Therefore, there
must exist a separable map $\Omega$ with two-qubit output, such that
the state $\Omega(\rho)$ satisfies condition (\ref{Cqubits}) for some
$(U,V,\theta)$, denote them by $(U_0,V_0,\theta_0)$.
Clearly, if $\Omega(\rho)$ satisfies (\ref{Cqubits}) there must exist
at least one value of $i$ in the Kraus decomposition of
(\ref{sepmap}) such that $(A_i\otimes B_i)~\rho~(A_i \otimes
B_i)^\dag$ also satisfies (\ref{Cqubits}). This implies that $\rho$
violates (\ref{C}) for $A= A_i^\dag U_0^\dag$, $B= B_i^\dag V_0^\dag$ and
$\theta=\theta_0$. This proves one direction of the lemma, let us show
the other.

Assume that $\rho$ violates (\ref{C}) for $(A_0,B_0,\theta_0)$. It is
straightforward to see that $\rho$ violates CHSH after stochastic
LOCC. Consider operation that transforms $\rho$ into $(A_0\otimes
B_0)^\dag\, \rho\, (A_0\otimes B_0)$. By assumption, the final state
satisfies (\ref{Cqubits}) with $U=V=\id$ and $\theta=\theta_0$, which
implies that it violates CHSH. $\Box$

\vs The above characterization is interesting on its own. Here we use it
to prove our main result.

\vs {\em Proof of the Theorem.} If $\sigma$ is separable then, $\rho
\in\c$ implies $\rho\otimes\sigma \in \c$. This is so because the
preprocessing by LOCC of $\rho$, before the Bell experiment, can
include the preparation of the state $\sigma$. Let us prove the other
direction of the theorem.

\begin{figure}[h!]
\scalebox{1}{\includegraphics{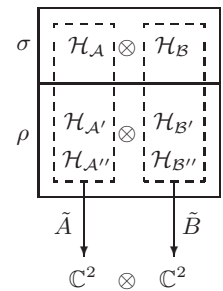}}
\caption{\label{Fig:Protocol} Schematic diagram illustrating
  the local filtering operations $\tilde{A}$ and $\tilde{B}$ involved
  in our protocol. The solid box on top is a schematic representation
  of the state $\sigma$ whereas that on the bottom is for the ancilla
  state $\rho$. Left and right dashed boxes, respectively, enclose the
  subsystems possess by the two experimenters $\A$ and $\B$.}
\end{figure}

>From now on $\sigma$ is an arbitrary entangled state acting on
$\H=\H_{\A} \otimes\H_{\B}$. Let us show that there always exists an ancilla
state $\rho\in\c$ such that $\rho\otimes\sigma \not\in \c$. Fix $\rho$
to act on the bipartite Hilbert space
$\left[\H_{\A'}\otimes\H_{\A''}\right] \otimes
\left[\H_{\B'}\otimes\H_{\B''}\right]$, where $\H_{\A'}=\H_\A$,
$\H_{\B'}=\H_{\B}$ and $\H_{\A''} =\H_{\B''} =\C^2$ (see Fig.~\ref{Fig:Protocol}).

Our aim is to prove that the state $\rho\otimes\sigma$ violates
(\ref{C}) for some choice of $A$, $B$, and $\theta$. In particular,
let
\begin{equation*}\label{tildeM}
  \tilde{A} = \ket{\Phi_{\A \A'}}\otimes
  \id_{\A''}\ ,\quad
  \tilde{B} = \ket{\Phi_{\B \B'}}\otimes
  \id_{\B''}\ , \quad \theta = \pi/4,
\end{equation*}
where $\ket{\Phi_{\A \A'}}$ is the maximally-entangled state between
the spaces $\H_{\A}$ and $\H_{\A'}$ (which have the same dimension),
and $\id_{\A''}$ is the identity matrix acting on $\C^2$ (analogously
for Bob). A little calculation shows that for any $\rho$
\begin{equation*}\label{eq}
    \tr\!\left[\rho\otimes\sigma\,
    (\tilde{A}\otimes\tilde{B}) H_{\pi/4}
    (\tilde{A}\otimes \tilde{B})^\dag\right]
    = \nu\, \tr\!\left[\rho\, (\sigma\t\otimes H_{\pi/4})\right]
\end{equation*}
where $\nu$ is a positive constant and $\sigma\t$ stands for the
transpose of $\sigma$. The inequality (\ref{C}) with $\theta=\pi/4$,
$A=\tilde{A}$, $B=\tilde{B}$ becomes
\begin{equation}\label{cond}
    \tr\left[\rho \left(\sigma\t
    \otimes H_{\pi/4}\right) \right]
    <0\ .
\end{equation}
For convenience, in the rest of the proof we allow $\rho$ to be
unnormalized. The only constraints on the matrices $\rho \in \c$ are:
positive semi-definiteness ($\rho \in \s$), and satisfiability of all
the inequalities (\ref{C}) in Lemma 1. $\c$ is now a convex cone, and
its dual cone is defined as
\begin{equation}\label{dual}
    \c^*=\{X : \tr[\rho\, X]\geq 0,\
    \forall \rho\in\c\}\ ,
\end{equation}
where $X$ are Hermitian matrices. Farkas' Lemma \cite{farkas} states
that all matrices in $\c^*$ can be written as non-negative linear
combinations of matrices $P \in \s$ and matrices $(A\otimes
B) H_\theta (A\otimes B)^\dag$ with $A:\C^2\rightarrow
\H_{\A'}\otimes\H_{\A''}$ and $B:\C^2\rightarrow
\H_{\B'} \otimes\H_{\B''}$.

We now show that there always exists $\rho\in\c$ satisfying
(\ref{cond}) by supposing otherwise and arriving at a contradiction.
Suppose that for all $\rho \in \c$ the inequality $\tr[\rho\,
(\sigma\t\otimes H_{\pi/4})]\geq 0$ holds, and thus the matrix
$\sigma\t\!\otimes H_{\pi/4}$ belongs to $\c^*$.  Applying Farkas'
Lemma \cite{farkas} we can write
\begin{equation}\label{pl}
    \sigma\t\! \otimes H_{\pi/4}=
    \int\!\! dx\, (A_x\otimes B_x) H_{\theta\!_x} (A_x\otimes
    B_x)^\dag + \int\!\! dy\, P_y\ ,
\end{equation}
which is equivalent to
\begin{equation}\label{principal0}
    \sigma\t\! \otimes H_{\pi/4} - \int\!\! dx
    \ \Omega_x\!\left( H_{\theta\!_x} \right) \geq 0\ ,
\end{equation}
where each $\Omega_x$ is a separable map (\ref{sepmap}). We prove in
Lemma 2 that (\ref{principal0}) requires that $\sigma$ is separable,
which gives the desired contradiction. Thus the result is proven.
$\Box$

In order to arrive at a contradiction from \eqref{principal0} it is
necessary to use the constraint that the maps $\Omega_x$ are
separable. The problem of characterizing the separable maps is hard in
general since it maps onto the separability problem for bipartite
states. However it turns out only to be necessary to determine the set
of separable maps that take Bell diagonal states to Bell diagonal
states and this can be done exactly~\cite{DLM}. This characterization
may be used to prove the following lemma and thereby our theorem.

\vs {\bf Lemma 2.} Let $\Omega_\theta: [\C^2] \otimes [\C^2]
\rightarrow [\H_{\A}\otimes \C^2]\otimes[\H_{\B} \otimes \C^2]$ be a family
of maps, separable with respect to the partition denoted by the
brackets.  Let $\mu$ be a unit-trace, positive semi-definite matrix
acting on $[\H_{\A}]\otimes[\H_{\B}]$ such that
\begin{equation}
\label{principal}
    \mu\t\! \otimes H_{\pi/4} - \int\!\! dx
    \ \Omega_x\!\left( H_{\theta_x} \right)
    \geq 0\ ,
\end{equation}
where $H_\theta$ is defined in (\ref{H}), then $\mu$ has to be
separable.

\vs {\em Proof.} Now, let us characterize the solutions
$\Omega_x$ of (\ref{principal}). The Bell basis is defined as
\begin{eqnarray}\label{bell}
  \ket{\Phi_{^1_2}} &=& 2^{-1/2}\left( \ket{00} \pm \ket{11} \right)\ , \\
  \ket{\Phi_{^3_4}} &=& 2^{-1/2}\left( \ket{01} \pm \ket{10} \right)\ .
\end{eqnarray}
The matrices $H_\theta$ are diagonal in this basis, $H_\theta=
\sum_{i=1}^4 N_\theta^i\, \Pi_i$, where $\Pi_i\equiv\proj{\Phi_i}$
are the Bell projectors and $N_\theta^i$ are the components of the vector
\begin{equation}\label{N}
    N_\theta = \left[
\begin{array}{c}
  1-\cos\theta-\sin\theta \\
  1+\cos\theta-\sin\theta \\
  1-\cos\theta+\sin\theta \\
  1+\cos\theta+\sin\theta \\
\end{array}
\right].
\end{equation}
For each value of $x$ define the sixteen matrices
\begin{equation}\label{omega}
    \omega_x^{ij} \equiv \tr_{\A'' \B''}\!\left[\left(\id\o\Pi_i\right)\,
    \Omega_x(\Pi_j)\right]\ ,
\end{equation}
for $i,j=1,2,3,4$, where the identity matrix $\id$ acts on
$\H_{\A'}\otimes\H_{\B'}$ and $\Pi_i$ acts on
$\H_{\A''}\otimes\H_{\B''}$. Each $\omega_x^{ij}$ is the result of a
physical operation, and hence positive. One can see $\omega_x^{ij}$ as
the Jamio{\l}kowski state corresponding to the map $\Omega_x$, after
``twirling" the input and output subsystems with the group of
unitaries that leaves Bell-diagonal states invariant. Projecting the
left hand side of \eqref{principal} onto the four Bell projectors
$\Pi_i$, and taking the relevant partial trace, we get
\begin{equation}\label{uk}
    \mu\t N_{\pi/4}^i - \int\!\! dx
    \sum_{j=1}^4\ \omega_x^{ij} N_{\theta\!_x}^j \geq 0
    \ ,
\end{equation}
for $i=1,2,3,4$. Denote by $M_x$
the $4\times4$ matrix with components $M_x^{ij}=\tr[\omega_x^{ij}]$.
Performing the trace on the left hand side of (\ref{uk}) we obtain the
four inequalities
\begin{equation}\label{io}
    N_{\pi/4} - \int\!\! dx
    \ M_x \cdot N_{\theta\!_x} \succeq \nv\ ,
\end{equation}
where $\nv$ is the 4-dimensional null vector, and the symbols $\cdot$
and $\succeq$ mean, respectively, standard matrix multiplication and
component-wise inequality. Consider the set of matrices $M$ that are
generated by tracing the left-hand side of (\ref{omega}) when
$\Omega_x$ is any separable map. The characterization of this set of
matrices is obtained in Ref.~\cite{DLM}, and goes as follows. Denote by
$\mathcal{D}$ the set of $4\times 4$ doubly-stochastic matrices, that
is, the convex hull of the permutation matrices
\cite{Bhat}. Denote by $\mathcal{G}$ the convex hull of all matrices
obtained when independently permuting the rows and columns of
\begin{equation}\label{e2}
G_0\equiv \left[
\begin{array}{cccc}
  1 & 1 & 0 & 0 \\
  1 & 1 & 0 & 0 \\
  0 & 0 & 0 & 0 \\
  0 & 0 & 0 & 0 \\
\end{array}
\right]\ .
\end{equation}
It is shown in Ref.~\cite{DLM} that, any matrix $M$ as defined above can be
written as
\begin{equation}\label{m}
    M=pD+qG\ ,
\end{equation}
where $D\in \mathcal{D}$, $G\in \mathcal{G}$, and $p,q\geq 0$. Then,
any solution of (\ref{io}) can be specified by giving
$(\theta_x,p_x,q_x,D_x,G_x)$. Using the fact that $G \cdot N_\theta
\succeq \nv$ for all $\theta$ and $G\in \mathcal{G}$, all solutions of
(\ref{io}) must satisfy
\begin{equation}\label{D}
    N_{\pi/4} \succeq \int\!\! dx
    \ p_x\, D_x \cdot N_{\theta\!_x}\ .
\end{equation}
Recall that this component-wise inequality entails four inequalities.
Adding them together we obtain the condition $\int dx \, p_x \leq 1$.
Now, denote by $\N$ the set of all vectors obtained by permuting the
components of $N_\theta$ (\ref{N}) when $\theta$ runs through
$[0,\pi/4]$. The convex hull of $\N$ ($\co~\N$) is precisely the set
of vectors that can be written as the right-hand side of (\ref{D})
under the constraint $\int dx \, p_x \leq 1$. The first inequality of
(\ref{D}) is
\begin{equation} \label{N1}
    1-\sqrt{2} \geq N^1\ ,
\end{equation}
where $N^1$ is the first component of $N\in \co~\N$. All vectors $N\in
\co~\N$ satisfy the inequality $1-\sqrt{2} \leq N^1$, and only
$N_{\pi/4}$ saturates it. Hence, the only possible value for the
right-hand side of (\ref{D}) is $N_{\pi/4}$. Substituting this into
\eqref{io}, and again using \eqref{m}, we obtain $-\int dx\ q_x\, G_x
\cdot N_{\theta\!_x} \succeq \nv$. But as said above, $G \cdot
N_{\theta} \succeq \nv$ for all $\theta$ and $G\in\mathcal{G}$, which
implies that for any solution $\int dx\ q_x\, G_x \cdot N_{\theta\!_x}
= \nv$. Therefore, inequality (\ref{io}) becomes $N_{\pi/4} - M_0
\cdot N_{\pi/4} \succeq \nv$, where $M_0$ is a doubly-stochastic
matrix such that $M_0\cdot N_{\pi/4}=N_{\pi/4}$.  Due to the form of
$N_{\pi/4}$, doubly-stochastic matrices that satisfy this equality
must have the form of
\begin{equation}\label{M_0} M_0 = \left[
\begin{array}{cccc}
  1 & 0 & 0 & 0 \\
  0 & 1- \eta & \eta & 0 \\
  0 & \eta & 1- \eta & 0 \\
  0 & 0 & 0 & 1 \\
\end{array}
\right]\ ,
\end{equation}
where $\eta\in [0,1]$.

The fact that each of the four components of the left-hand side of
(\ref{io}) is zero, implies that the left-hand side of (\ref{uk}) is
traceless for all $i$. The only positive matrix with zero trace is the
null matrix, therefore
\begin{equation}\label{fin}
    \mu\t N_{\pi/4}^i =
    \sum_{j=1}^4\, \omega_0^{ij} N_{\pi/4}^j,\quad i=1,2,3,4,
\end{equation}
where $\omega_0$ is any $\omega_x$ that gives rise to $M_0$. Using the
same argument, the pairs $(i,j)$ for which $M_0^{ij}=0$ are such that
$\omega_0^{ij}=0$. Therefore, by adding the equalities in (\ref{fin})
corresponding to $i=2,3$, and using the definition of $\omega_0^{ij}$
in \eqref{omega}, we obtain
\begin{equation}\label{fin2}
    2\, \mu\t = \tr_{\A'' \B''}\!\left[\left(\id\o\Psi\right)
    \Omega_0(\Psi)\right]\ ,
\end{equation}
where $\Psi=\Pi_2+ \Pi_3$, and $\Omega_0$ is any $\Omega_x$ that
gives rise to $\omega_0$. Using the PPT criterion \cite{Hppt} one
can check that the (unnormalized) two-qubit state $\Psi$ is a
separable state. Equation (\ref{fin2}) implies that $\mu\t$ is the
output of a separable map applied to a separable input state, and
hence is a separable state as we wanted to prove. $\Box$

\vs{\bf Summarizing}, for each entangled state $\sigma$ there exists a
protocol (involving the state $\rho$ associated with the theorem) in
which $\sigma$ cannot be substituted by an arbitrarily large amount of
classical correlations, without changing the result:

\vs {\em Entangled states are the ones that cannot be simulated by
  classical correlations.}

\vs This provides us with a new interpretation of entanglement in
terms of the behavior of the states, in contrast with the usual
definition in terms of the preparation of the states.

\vs Differently, one can be interested in the set of bipartite states
$\sigma$ which do not admit a LHVM description in scenarios where
no other kind entanglement is present. That is, $\sigma$ may be
processed with more copies of itself $\sigma^{\otimes n}$ but never
with different entangled states $\rho$. Following \cite{Masymp} we say
that a state $\sigma$ violates a Bell inequality asymptotically, if
after jointly processing by LOCC a sufficiently large number of copies
of a state, the result violates the Bell inequality. In Ref.~\cite{Masymp}
it is proven that the states violating CHSH asymptotically are the
distillable ones. This, together with the results of the present
paper, establishes an appealing picture:
\begin{eqnarray*}
  \mbox{entangled} &\Longleftrightarrow& \mbox{non-simulable} \\
  \mbox{distillable} &\Longleftrightarrow&
  \mbox{asymptotic violation of CHSH}
\end{eqnarray*}
\\

Entangled states are the ones that cannot be generated
from classical correlations plus local quantum operations.  We show
that in the bipartite case, one can equivalently define entangled
states as the ones that cannot be simulated by classical correlations.\\

\vs {\bf Acknowledgments.} The authors are thankful to Fernando
Brand\~ao and Nick Jones for useful comments. This work is supported
by the EU Project QAP (IST-3-015848) and the Australian Research Council.

\end{document}